\documentclass[aps,prl,twocolumn]{revtex4-1}
\usepackage{amsmath,bm,epsfig,,subfigure}
\usepackage{color}
\usepackage[normalem]{ulem}

\def\Fbox#1{\vskip1ex\hbox to 8.5cm{\hfil\fboxsep0.3cm\fbox{%
  \parbox{8.0cm}{#1}}\hfil}\vskip1ex\noindent}  

\newcommand{\B}[1]{{\bm{#1}}}
\newcommand{\C}[1]{{\mathcal{#1}}}    
\let \= \equiv \let\*\cdot \let\~\widetilde \let\-\overline

\begin{document}
\title{Micro-alloying and the Toughness of Glasses: Modeling with Pinned Particles}
\author{Ratul Dasgupta$^1$, Pankaj Mishra$^1$,  Itamar Procaccia$^1$ and Konrad Samwer$^2$}
\affiliation{$^1$Department of Chemical Physics, The Weizmann
 Institute of Science, Rehovot 76100, Israel\\ $^2$I.Physikalisches Institut,
Universitaet G\"ottingen, Friedrich-Hund-Platz 137077  G\"ottingen, Germany.}
\date{\today}
\begin{abstract}

The usefulness of glasses, and particularly of metallic glasses, in technological applications is often limited
by their toughness, which is defined as the area under the stress vs. strain curve before plastic yielding.
Recently toughness was found to increase significantly by the addition of small concentrations of foreign
atoms that act as pinning centers. We model this phenomenon at zero temperature and quasi-static straining with randomly positioned particles that participate
in the elastic deformation but are pinned in the non-affine return to mechanical equilibrium. We find a very strong
effect on toughness via the increase of both the shear modulus and the yield stress as a function of the density of pinned particles. Understanding the results calls for analyzing separately the elastic, or ``Born term" and the contributions of the ``excess modes" that result from glassy disorder. Finally we present a scaling theory that collapses the data on one universal curve as a function of rescaled variables.

\end{abstract}

\maketitle

{\bf Introduction}: In the field of metallic glasses an old but very important protocol called ``micro-alloying" is getting recently very high attention. The protocol comprises the addition of foreign atoms with concentration of the
 order of 1\% to an otherwise regular metallic glass. Such micro-alloying results in dramatic changes in the physical and chemical properties of the material,  including its  glass forming ability \cite{07Wan}, diffusion properties \cite{09CDES}, corrosion resistance \cite{07Wan} and plastic properties \cite{11DLGSHJR}. These colossal changes,which are found even in the sub-percent alloying range as demonstrated by Ref. \cite{11DFCSGJ} for 0.5\% Ag additions require a different approach compared to the normal alloying of two or more components in the middle of the phase diagram. Changes in the electronic or chemical structure of the global composition are not sufficient to explain  the measured large effects by such minor additions. This is also clearly shown in recent toughness measurements \cite{12GDCJ}, where the addition of 1-2\% of Si/Sn changes the notch toughness by 50\%.
 
In an earlier work \cite{09CDES} it was suggested that the medium range order of the metallic glasses is reorganized and the stress distribution of that reconfiguration \cite{07HDJS} causes the change in long range property. In order to verify this suggestion we studied in a most recent MD-simulation \cite{11CKPS} the influence of pinned particles on the ``global" property of the metallic alloy. Pinning particles in a very small concentration range act like too ``big" or too ``small" particles that are introduced in a nearly perfect matrix of a glassy system. We found that the $\beta$-process of the amorphous system, which is well known to be responsible for aging and diffusion in the glassy state below the glass transition \cite{12YSWW} is totally suppressed by a small amount (up to 5\%) of pinned particles. This is far below the percolation limit of the system. In this Letter we show that the pinning protocol might provide a general approach to understand the effects of micro-alloying in bulk metallic 
glasses because it also affects the yield stress, the modulus and the toughness. Predictions are made below for further comparison with experiments.

{\bf Numerical simulations}: To prepare accurate data for the present discussion
we have performed 2-dimensional Molecular Dynamics simulations on a standard binary system which is known to be a good glass former. We employed a $50-50$ binary Lenard-Jones mixture with a potential energy for a pair of particles labeled  $i$ and $j$ in the form
\begin{eqnarray}
&&U_{ij}(r_{ij}) = 4\epsilon_{ij}\Big[\Big(\frac{\sigma_{ij}}{r_{ij}}\Big)^{12} - \Big(\frac{\sigma_{ij}}{r_{ij}}\Big)^{6} + A_0 + A_1\Big(\frac{r_{ij}}{\sigma_{ij}}\Big)\nonumber\\&&+ A_2\Big(\frac{r_{ij}}{\sigma_{ij}}\Big)^2\Big] \ ,
\end{eqnarray}
where the parameters $A_0$, $A_1$ and $A_2$ are added to smoothen the potential at a scaled cut-off of $r/\sigma = 2.5$ upto the second derivative. The parameters $\sigma_{_{AA}}$, $\sigma_{_{BB}}$ and $\sigma_{_{AB}}$ were chosen to be $2\sin(\pi/10)$, $2\sin(\pi/5)$ and $1$ respectively and $\epsilon_{_{AA}} = \epsilon_{_{BB}} = 0.5, \epsilon_{_{AB}} = 1$(see \cite{Langer1998}). The particle masses were taken to be equal. The samples were  prepared using high-temperature equilibration followed by a quench to zero temperature ($T=0.001$) (see \cite{Falk2007}). Our quench rate was $3.2\times 10^{-6}$ in LJ units.  All our procedures are
performed with a fixed total number of particles $N$ and a constant volume such that the overall particles
density $\rho$ is strictly constant and equals $0.976$.
\begin{figure}
\includegraphics[scale = 0.45]{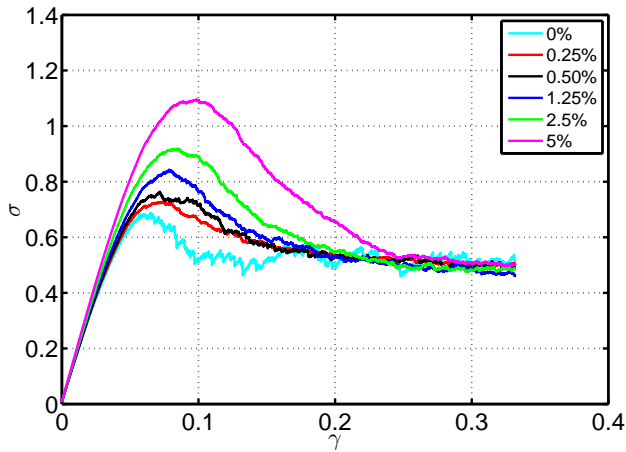}
\caption{(Color Online). Typical stress vs. strain curve obtained for AQS
straining of one realization of a system of $N$=20000 particles at different densities $\rho_{_P}$ of pinned particles. Note the significant change in the shear modulus (the slope at $\gamma=0$)
and of the yield stress where the system yields to plastic flow.}
\label{sigvsgam}
\end{figure}

Having prepared the amorphous solid in very much the standard way developed in the field, we now depart from
the usual procedure by selecting randomly a set $\{P\}$ of $P$ particles that will be referred to below as the ``pinned particles". These are chosen such that their density is $\rho_{_P}\ll \rho_{_0}$, where $\rho_{_P}$ and $\rho_{_0}$ are, respectively, the density of pinned and not pinned particles ($\rho=\rho_{_P}+\rho_{_0}$). At this point we strain our
 sample using an athermal quasi-static (AQS) protocol to examine
their stress vs. strain curves. In this procedure the particle positions in the system are first
changed by the affine transformation
\begin{equation}
x_i\to x_i+\delta \gamma y_i; \quad y_i\to y_i \ .
\end{equation}
During this affine step the pinned particles are allowed to participate like all the other particles.
This transformation results in the system not being in mechanical equilibrium, and we therefore allow
the non affine transformation $\B r_i\to \B r_i+\B u_i$ which annuls the forces between the particles,
returning the system to mechanical equilibrium. In this step the pinned particles are not allowed to
participate; they are held fixed at the positions that they attained at the end of the affine step.
The other particles are moving, directed by a gradient energy minimization, to seek new positions that
annul the forces between all the particles, pinned and not pinned. In other words during the energy minimization
we add counter force such that the net force experienced by the pinned particles is constrained to be zero. Thus the main effect of the pinned
particles is to constrain the system not to follow the standard conjugate gradient path but to find new ``inherent states" or local minima in the energy landscapes,
different from those that could be found if all the particles were allowed to move in the non-affine step.

\begin{figure}
\includegraphics[scale = 0.45]{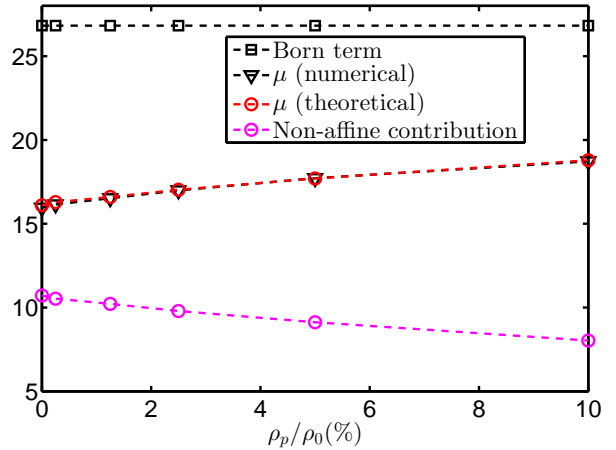}
\caption{(Color Online). Shear modulus as a function of the density $\rho_{_P}/\rho_0$ of pinned particles (N = 20000). In black inverted
triangles we show $\mu$ as measured from the slope of Fig. \ref{sigvsgam} at $\gamma=0$. In red dots we show $\mu$ as
computed from the theory. The Born contribution is given by the black squares, showing an almost independence of $\rho_{_P}$.
The non-affine contribution is shown in terms of the magenta circles. Of course the shear modulus itself
is the difference between the Born term and the non-affine contribution.}
\label{mu}
\end{figure}

In Fig.~\ref{sigvsgam} we show the raw data of stress vs strain for this material
with varying densities of pinned particles as presented in the inset. Shown are averages over 5 realizations
of the glass, with the same configuration of randomly chosen pinned particles.
We observe that both the shear modulus and the yield stress (where the system yields to plastic flow) increase significantly when the density of pinned particles $\rho_{_P}$ is increased. In Fig. \ref{mu} we present the shear modulus as a function of $\rho_{_P}$ (see the middle curve with inverted triangles). In the rest of the Letter we want to clarify this phenomenon in a quantitative way.

{\bf Theory}: By definition the shear modulus is the second derivative of the energy of the system
with respect to the strain $\gamma$, i.e.
\begin{equation}
\mu = \frac{1}{V}\frac{d^2 U(\B r_1,\cdots,\B r_N;\gamma)}{d \gamma^2} \ .
\end{equation}
In our process the full derivative with respect to $\gamma$ translates to two contribution, one the direct
partial derivative with respect to $\gamma$ and the other, via the chain rule, the contribution due to the non-affine part of the transformation:
\begin{equation}
\frac{d}{d\gamma} = \frac{\partial}{\partial \gamma} + \sum_{i\notin \{P\}}\frac{\partial}{\partial \B u_i} \cdot \frac{\partial \B u_i}{\partial \gamma} \equiv  \frac{\partial}{\partial \gamma} +  \sum_{i\notin \{P\}}\frac{\partial}{\partial \B r_i} \cdot \frac{\partial \B u_i}{\partial \gamma}\ ,
\end{equation}
where the second equality follows from the form of the non-affine transformation where $d\B r_i=d\B u_i$.
To understand how to compute the constrained sum we need to recall that the force $\B f_i\equiv -\partial U/\partial \B r_i$ is
zero before and after the affine and non-affine steps. Thus
\begin{equation}
-\frac{d}{d\gamma} \frac{\partial U}{\partial \B r_i} = 0= -\frac{\partial^2 U}{\partial \gamma\partial \B r_i} - \sum_{j\notin \{P\}}
\frac{\partial^2 U}{\partial \B r_i\partial \B r_j} \cdot \frac{\partial \B u_j}{\partial \gamma} \ .
\label{straightu}
\end{equation}
Denote now as usual the Hessian matrix $\B H$ and the non affine ``force" $\B \Xi$ \cite{10KLP}:
\begin{equation}
H_{ij} \equiv \frac{\partial^2 U(\B r_1,\cdots,\B r_N;\gamma)}{\partial \B r_i \partial \B r_j}\ , \quad
\Xi_i \equiv \frac{\partial^2 U(\B r_1,\cdots,\B r_N)}{\partial \gamma\partial \B r_i } \ .
\end{equation}
Inverting Eq. (\ref{straightu}) we find
\begin{equation}
\frac{d\B u_i}{d\gamma}\Big|_{i\notin \{P\}} = -\sum_{i\notin \{P\}} H^{-1}_{ij}\cdot \Xi_j \ .
\end{equation}

Applying these results to the definition of $\mu$ we end up with the exact expression
\begin{equation}
\mu=\frac{1}{V}\frac{\partial^2 U(\B r_1,\cdots,\B r_N;\gamma)}{\partial \gamma^2}-\frac{1}{V}
\sum_{i\notin \{P\}}\Xi_i\cdot  H_{ij}^{-1} \cdot \Xi_j \ , \label{defmu}
\end{equation}
where the first term is the well known Born contribution which we denote below as $\mu_B$. The second term exists only due to the non-affine
displacement $\B u_i$ and it includes an unusual matrix of rank $(N-P)\times N$.
Needless to say, before we compute the non-affine contribution in Eq.~(\ref{defmu}) we need to remove the two Goldstone modes
with $\lambda=0$ which are the result of translation symmetry.

It is very important to stress at this point that the separation between the Born term and the
non-affine term is not an arbitrary one. The Born term is very insensitive to pinned particles in our
example, and this is usually the case: it is only sensitive to average properties like density,
average number of neighbors and interactions \cite{65Zwa}. In Fig.~\ref{mu} we show the result of calculating
the Born term for all our samples as a function of the density of pinned particles, (see black squares in Fig.~\ref{mu}), and there is only minor dependence. This is not the case for the non-affine term, whose direct calculation is also shown in the same figure in magenta circles. We see that this term changes significantly, taking upon itself the full blame of the change in the shear modulus as a function of the density of pinned particles. The difference of the two terms agrees to very high accuracy with the direct measurement of the shear modulus from the slopes of the curves in Fig.~\ref{sigvsgam} at $\gamma=0$. We note that in other cases in which, say, the shear modulus
depended on the quench rate from the melt, the physics was determined by the density of states which varied considerably
with the quench rate \cite{13ABP}. This is NOT the case here. The full Hessian matrix is oblivious of the pinned particles,
and its spectrum remains invariant when we change their density. The full reason for the change in shear modulus is embodied in the
restricted sum in Eq.~(\ref{defmu}), respecting the fact that the pinned particles are not involved in the non-affine
step of our straining protocol.
\begin{figure}
\includegraphics[scale = 0.45]{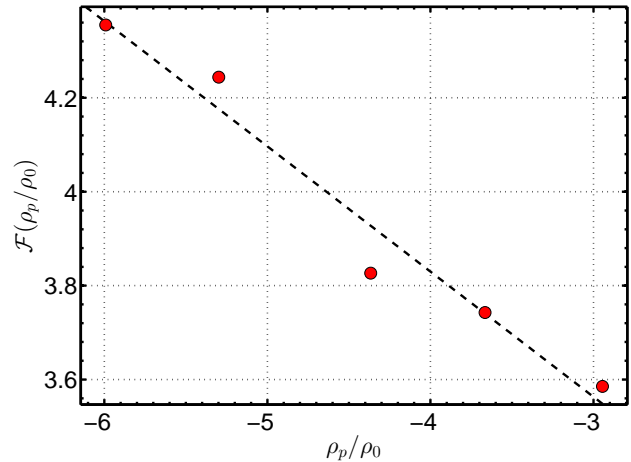}
\caption{(Color Online). Log-log plot of the rescaled shear modulus $\mu(\rho_{_{P}}) - \mu(\rho_{_{P}}=0)/({\epsilon_{_{AB}}}{\rho_{_P}})$ as a function of the dimensionless ratio
 $\rho_{_{P}}/\rho_0$. The slope is $\alpha=-0.25\pm 0.02$ and the data agree with the scaling function ansatz as represented in Eq. (\ref{scfun}).}
\label{scalemu}
\end{figure}

{\bf Scaling theory}:
Having at hand the exact microscopic theory to evaluate the effect of the pinned particles on the shear modulus,
we turn now to a scaling approach which may yield the phenomenological dependence of $\mu$ on $\rho_{_{P}}$. We
start with the scaling ansatz
\begin{equation}
\mu(\rho_{_{P}}) - \mu(\rho_{_{P}}=0) = {\epsilon_{_{AB}}} {\rho_{_P}} \C F(\rho_{_{P}}/\rho_0) \ ,
\label{scmu}
\end{equation}
where the product ${\epsilon_{_{AB}}} {\rho_{_P}}$ exhausts the dimensionality of the LHS and $\C F$ is
a scaling function of a dimensionless variable. Expecting that the scaling function might be represented by
a power law we plot in a log-log plot $\mu(\rho_{_{P}}) - \mu(\rho_{_{P}}=0)/({\epsilon_{_{AB}}}{\rho_{_P}})$ as a function of
$\rho_{_{P}}/\rho_0$. This plot is shown in Fig.~\ref{scalemu}, indicating that to a very good approximation
the scaling function $\C F$ is indeed a power law,
\begin{equation}
\C F(\rho_{_{P}}/\rho_0) \propto (\rho_{_{P}}/\rho_0)^\alpha , \quad \alpha\approx -0.25\pm 0.02 \ .
\label{scfun}
\end{equation}
collecting these results together indicates a phenomenological law for the shear modulus in the form
\begin{equation}
\mu(\rho_{_{P}}) - \mu(\rho_{_{P}}=0) \propto {\rho_{_P}}^{0.75} \ .
\end{equation}
It should be very interesting to compare this theoretical prediction to experiments on micro-alloying.

\begin{figure}
\includegraphics[scale = 0.45]{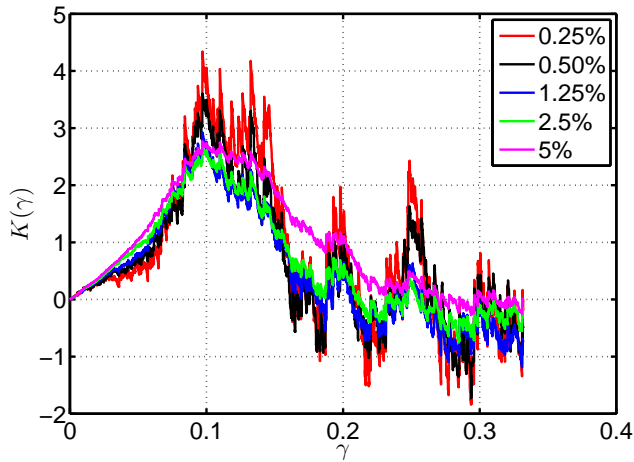}
\caption{(Color Online). The universal function $K(\gamma)$ which is obtained as explained in the text.
This scaling function represents a collapse of all the data that appear in Fig.~\ref{sigvsgam}. Note that
in this collapsed form the values of $\sigma_{_Y}$ as well as of $\gamma_{_Y}$ become independent of
the pinning.}
\label{fullscale}
\end{figure}

Encouraged by this result we proceed further to seek a scaling form for the whole stress vs strain response
as a function of rescaled variables with the aim of providing scaling collapse for the data in Fig.~\ref{sigvsgam}.
We start with the obvious scaling ansatz
\begin{equation}
\sigma(\gamma,\rho_{_{P}}) - \sigma(\gamma,\rho_{_{P}}=0) = {\epsilon_{_{AB}}} {\rho_{_P}} \tilde{\C G}(\gamma,\rho_{_{P}}/\rho_0) \ .
\end{equation}
From the analysis of scaling of the shear modulus we conclude that
\begin{equation}
\lim_{\gamma\to 0} \tilde{\C G}(\gamma,\rho_{_{P}}/\rho_0) \propto {\rho_{_P}}^{-0.25} \ .
\end{equation}
The exponent in the scaling law changes when $\gamma$ and $\sigma$ increase.  We analyzed the data in the region of $\gamma\approx \gamma_{_Y}$ and found that 
\begin{equation}
\lim_{\gamma\to \gamma_{_Y} } \tilde{\C G}(\gamma,\rho_{_{P}}/\rho_0) \propto {\rho_{_P}}^{-0.50} \ .
\end{equation}
We can combine the last two equations into an interpolation formula which provides the global scaling
form of $\tilde {\C G}$. Defining $\sigma_{_Y}$ as the value of $\sigma$ at which the stress reaches a maximum in Fig~\ref{sigvsgam}, we write
\begin{equation}
\frac{\sigma(\gamma,\rho_{_{P}}) - \sigma(\gamma,\rho_{_{P}}=0)}{{\epsilon_{_{AB}}} \rho_{_P}} =  K(\gamma)
\frac{\sigma_{_Y} (\rho_{_P}/\rho_0)^{-0.5}}{\sigma+(\sigma_{_Y}-\sigma)(\rho_{_P}/\rho_0)^{-0.25}}  \ ,
\label{ugly}
\end{equation}
where $K(\gamma)$ is expected to be a universal scaling function of $\gamma$, independent of concentration
of pinned particles. Denoting
\begin{equation}
\C G(\sigma, \rho_{_{P}}/\rho_0) \equiv \frac{\sigma_{_Y} (\rho_{_P}/\rho_0)^{-0.5}}{\sigma+(\sigma_{_Y}-\sigma)(\rho_{_P}/\rho_0)^{-0.25}} \ ,
\end{equation}
we present in Fig.~\ref{fullscale} the LHS of Eq. (\ref{ugly}) divided by $\C G$ as a function of $\gamma$.
The analysis indicates that the resulting plot should be the universal function $K(\gamma)$.

 {\bf Conclusions}: We have demonstrated that our model micro-alloying which is realized by pinning randomly a small percentage of particles during the non-affine relaxation step has a huge effect on the mechanical properties, i.e., shear modulus, yield-stress and toughness. For example the yield stress changes by more than $50\%$ upon increasing the concentration of the pinned particles by $5\%$. The effect on the shear modulus had been explained by an exact theory separating the Born term from the non-affine contribution. It was shown that the latter takes full responsibility for the changes in shear modulus. The exact theory matched perfectly with the simulation results. Probably the most important result of this letter is the offered scaling theory for {\it both} the shear modulus and the entire stress vs. strain curves. This theory allows for a data collapse summarizing large amount of data in terms of a few simple functions of dimensional variables. It would be extremely interesting to test the applicability of similar scaling ideas to experimental examples of the effect of micro-alloying.

Acknowledgements: this work was supported by the Israel Science Foundation, the German-Israeli Foundation and
by the ERC under the STANPAS ``ideas" grant. We thank Oleg Gendelman for useful discussions.

\end{document}